\documentclass[12pt]{article}
\usepackage{enumerate}
\usepackage{subfigure}
\usepackage{amsmath}
\usepackage{amssymb}
\usepackage{amsthm} 
\usepackage{psfrag}
\usepackage{graphicx}
\usepackage[numbers,sort&compress]{natbib}

\newcommand{\be}{\begin{equation}}
\newcommand{\ee}{\end{equation}}
\newcommand{\bel}[1]{\begin{equation}\label{#1}}
\newcommand{\bea}{\begin{eqnarray}}
\newcommand{\eea}{\end{eqnarray}}
\newcommand{\ba}{\begin{array}}
\newcommand{\ea}{\end{array}}

\newtheorem{thm}{Theorem}
\newtheorem*{cor}{Corollary}
\theoremstyle{remark}
\newtheorem*{remark}{Remark}

\DeclareMathOperator{\Prob}{Prob}

\begin{document}

\title{Bethe ansatz and current distribution for the TASEP with 
particle-dependent hopping rates}

\author{A.~R\'akos\footnote{Present address: \it Department of Physics 
of Complex Systems, Weizmann Institute of Science, Rehovot 76100, Israel}, 
G.~M.~Sch\"{u}tz\\
\it \small Institut f\"{u}r Festk\"{o}rperforschung, Forschungszentrum
J\"{u}lich, \\ 
\it \small 52425 J\"{u}lich, Germany.}
\date{June 20, 2005}

\maketitle

\begin{abstract}
Using the Bethe ansatz we obtain in a determinant form the exact solution of the 
master equation for the conditional probabilities of the totally asymmetric
exclusion process with particle-dependent hopping rates on $\mathbb{Z}$. From this
we derive a determinant expression for the time-integrated current for
a step-function initial state.
\end{abstract}

\section{Introduction}

In the usual one-dimensional totally asymmetric simple exclusion process 
(TASEP) particles hop randomly in continous time on the integer lattice 
$\mathbb{Z}$. A hopping event occurs independently for each particle after an 
exponentially distributed random time with parameter 1 to its right 
neighboring site, provided this site is empty; otherwise the attempted move 
is rejected \cite{Ligg99,Schu01}. This is the exclusion interaction
between biased random walkers, which 
physically models a short-range hard-core repulsion of identical 
driven diffusive particles.
Here we consider the case of distinct particles where each particle 
$i$ has its own intrinsic hopping rate $v_i$.

This model has been studied earlier by other authors. To our knowledge
it was first introduced (with a random choice of rates $v_i$, drawn
from some given distribution) by
Benjamini et al.\ \cite{Benj96} who proved the existence of a product
invariant distribution (as seen from a tagged particle) above some
critical density $\rho^\ast$ of particles. They also proved for this case the 
hydrodynamic limit of an associated zero-range process (ZRP) which 
is obtained from the TASEP by identifying particles with the sites
of a new 1-D lattice and the interparticle distance (number of empty
sites between particles $i,i+1$) as occupation number at site $i$
of that lattice. Subsequently it was shown for periodic boundary
conditions that below the critical density $\rho^\ast$ a macroscopic traffic 
jam, trailing the slowest particle, develops \cite{Evan96,Krug96}.
The macroscopic dynamical signature of the transition under Eulerian scaling
is a discontinuity in the coarse-grained density at the front of
a rarefaction wave that develops out of an initial step function 
profile \cite{Sepp99}. The stationary distribution of an open system
where particles are injected and extracted was obtained numerically 
\cite{Beng99}. Somewhat surprisingly, the theory of
boundary-induced phase transitions developed by Kolomeisky et al.\ 
\cite{Kolo98} accounts for the stationary phase diagram of the system
as a function of the effective boundary densities.

The jamming transition occurring in this model is a classical analog of 
Bose-Einstein condensation and one of the
motivations to study this system in more detail. An analogous transition
occurs also if the rates are identical for each particle, but dependent
on the lattice distance to the
next particle, see \cite{Hann05} for a recent
review (in terms of the ZRP) and 
\cite{Gros03,Godr03,Levi04,Harr05,Kaup05,Godr05} 
for the current developments. A second motivation comes from the recent
interest in particle systems with several conservation laws \cite{Schu03}.
By choosing the $v_i$ in some way out of $p$ distinct values the
model indeed becomes a particle system with $p$ conservation laws
where, unlike in a similar class of models introduced by Karimipour
\cite{Kari99a,Kari99b,Khor00}, no overtaking of particles is permitted. 
Exact results
for such multi-component systems are scarce. Our contribution aims at
providing some tools for future exact and rigorous analysis. 
We first solve the master equation for an $N$-particle system
defined on $\mathbb{Z}$ with the Bethe ansatz generalizing the approach
taken by one of us earlier \cite{Schu97}. This yields a determinant
formula for the conditional probabilities which we then use to derive
an exact expression for the distribution of the time-integrated
current for the step initial configuration envisaged in \cite{Sepp99}. Our 
result generalizes that obtained recently for the TASEP with constant
rates \cite{Sasa04,Rako05}.

\section{Solution of the master equation}

We consider initial states such that there is no particle beyond
a lattice point $x_\text{max} < \infty$ at time $t=0$. Since the process is 
totally asymmetric and defined on $\mathbb{Z}$ the $N$ rightmost particles are
not influenced by particles to their left. Without loss of generality
we may therefore consider $N$-particle initial states where the
particles are labelled from 1 through N from left to right.
Therefore at any finite time time $t\geq 0$ a configuration of the system can 
be represented by an increasing sequence of integer numbers 
($x_1, x_2, \cdots, x_N$), $x_i$ denoting the position of the $i$th particle with hopping rate $v_i$. We assume $0 < v_i \leq v_\text{max}$
where  $v_\text{max} < \infty$ sets the microscopic time scale. We remark that allowing for $v_\text{max}=\infty$ is 
equivalent to considering particles covering more than 1 (neighboring)
site \cite{Alca99,Laka03,Shaw03,Alca03,Scho04,Scho05}. Such a system can be 
mapped to an exclusion process with standard particles (of lattice size 1) by 
an appropriate coordinate shift. The case $v_i =0$ is trivial in so far as this 
amounts to cutting the lattice $\mathbb{Z}$ into finite segments in which after some 
time all particles pile up at the right boundary of the segment and the 
dynamics freeze.

In case of equal hopping rates the exchange of particles is irrelevant as they
are indistinguishable. However, in this generalized version one has to define 
the corresponding exchange rules separately since the particles have their own 
identity. In our model particles cannot exchange so one can interpret it as a 
one-lane traffic model with cars having their own preferred speed. 

The master equation for the probability of finding particles 
at sites $x_i$ is the following (the $t$-dependence is dropped for
simplicity of notation):
\begin{multline}
\label{me1}
\frac{d}{dt} P(x_1, x_2, \cdots, x_N) = v_1 P(x_1-1, x_2, \cdots x_N) \\
+ v_2 (1-\delta_{x_2-1, x_1}) P(x_1, x_2-1, \dots, x_ N) \\ 
\cdots + v_N (1-\delta_{x_N-1, x_{N-1}}) P(x_1, x_2-1, \dots, x_N-1) \\
-(v_1(1-\delta_{x_2-1, x_1})+v_2(1-\delta_{x_3-1, x_2})+ \cdots + v_N )
P(x_1, x_2, \dots, x_N),
\end{multline}
where $P(x_1, x_2, \cdots, x_N)$ is the probability of the configuration 
($x_1, x_2, \cdots, x_N$). In the spirit of \cite{Schu97} 
we extend the range of definition of $P(x_1, x_2, \cdots, x_N)$
from the physical domain $x_i < x_{i+1}$
to $\mathbb{Z}^N$.
By requiring the boundary condition
\begin{multline}
\label{bc}
v_{k+1} P(x_1, \cdots, x_{k-1}, x_k, x_k, x_{k+2}, \cdots, x_N) \\ = {v_k} 
P(x_1, \cdots, x_{k-1}, x_k, x_k+1, x_{k+2}, \cdots, x_N)
\end{multline}
to be valid for all $t > 0$
the master equation (\ref{me1}) reduces to the simple form
\begin{multline}
\label{me2}
\frac{d}{dt} P(x_1, x_2, \cdots, x_N) = \sum_{i=1}^N v_i 
P(x_1, \dots, x_{i-1}, x_i-1, x_{i+1}, \cdots, x_N)  \\ 
-\sum_{i=1}^N v_i P(x_1, x_2, \dots, x_N) \equiv -\hat H 
P(x_1, x_2, \cdots, x_N).
\end{multline}
The second equality defines the linear operator $\hat H$ generating the
time evolution of the probability distribution. For given initial configuration
$(y_1, y_2, \cdots, y_N)$ this quantity is the conditional probability
$P(x_1, \cdots, x_N|y_1, \cdots, y_N;t)$.

It is obvious that (\ref{me2}) can be solved by Fourier transformation
with a ``momentum'' variable $k_i$ associated with each particle coordinate
$x_i$. However, a straightforward Fourier ansatz does not satisfy the boundary
condition (\ref{bc}). This problem can be resolved by noting
that there is some freedom by choosing
a linear combination of Fourier transforms over all permutations of the 
dummy momentum variables $k_i$ with complex coefficients (wave amplitudes) 
that depend on the $k_i$. It is the essence of the Bethe ansatz to take
this approach and to implement it by factorizing the wave amplitudes for
the $N$-particle problem into amplitudes for the two-particle problem.
It is done such that for each permutation of the momentum variables the 
$N$-particle amplitude picks up the corresponding two-particle amplitude as 
extra factor.

In the present case this strategy also turns out to be successful, 
but requires
some modification due to the occurrence of the different hopping rates $v_i$.
In order to obtain the eigenvalues and eigenfunctions of the ``Hamiltonian''  
$\hat H$ with the boundary condition (\ref{bc}) we write
\begin{equation}
\frac{d}{dt} P_\mathbf{k} (x_1, \cdots, x_N) = -E_\mathbf{k} 
P_\mathbf{k} (x_1, \cdots, x_N).
\end{equation}
where $\mathbf{k}$ represents the conjugate momenta $k_1,\dots,k_N$.
Modifying the Bethe ansatz to account for the $v_i$ leads to eigenfunctions
of the form \cite{Kari99b}
\begin{equation} 
\label{ef1}
P_{k_1, \cdots, k_N}(x_1, \cdots, x_N) = 
\prod_{i=1}^N {v_i}^{x_i} \sum_{\pi} S_{\pi} 
\exp\left(i \sum_{j=1}^N x_j k_{\pi(j)}\right),
\end{equation}
where the sum is taken over all permutations $\pi$ of $(1,2,3,\cdots,N)$. The 
above eigenfunctions correspond to the eigenvalues 
\begin{equation}
E_{k_1, \cdots, k_N} = \sum_{j=1}^N \left( v_j - e^{-ik_j} \right).
\end{equation}
With the inverse Fourier transformation
\be
\label{invfour}
P(x_1,\cdots, x_N) = \prod_{i=1}^N \int \frac{dk_i}{2\pi} 
\mbox{e}^{-E_\mathbf{k}t}
P_{\mathbf{k}}(x_1,\cdots, x_N)
\ee
it is trivial to show that this ansatz satisfies (\ref{me2}) with arbitrary wave amplitudes $S_\pi(\mathbf{k})$ .

On the other hand from the boundary condition (\ref{bc}) it follows that
\begin{equation}
\label{S}
S_{T_k \pi} = -\frac{1-v_{k}e^{ik_{\pi(k+1)}}}{1-v_{k}e^{ik_{\pi(k)}}} 
S_\pi,
\end{equation}
where the operator $T_k$ changes the $k$th and $k+1$st element of $\pi$
and $S_{(1,2,\dots,N)}$ is an arbitrary function defined by the
initial condition to be satisfied at $t=0$, see next section. 

In order to obtain the $N$-particle amplitude $S_\pi$ for a given permutation $\pi$ one has to decompose this permutation into a sequence of nearest neighbour exchanges and take the product of factors appearing in (\ref{S}). Although this decomposition is not unique the resulting $S_\pi$ is well-defined if certain relations, known as Yang-Baxter equations, are satisfied. Instead of checking this explicitly one can easily show that the r.h.s.\ of (\ref{ef1}) with an $S_\pi$ satisfying (\ref{S}) can be written in the following determinant form
\begin{multline}\label{det}
P_{k_1, \cdots, k_N}(x_1, \cdots, x_N) \\ = S_{(1,\dots,N)}
\prod_{i=1}^N {v_i}^{x_i} \det \left[ \prod_{j=1}^{n-1} (1-v_j e^{ik_m})^{-1} 
\prod_{j=1}^{m-1} (1-v_j e^{ik_m}) e^{i k_m x_n} \right]_{m,n}.
\end{multline}
This verifies uniqueness by construction and therefore also implies the Yang-Baxter equations.

\section{Conditional probability}

In order to obtain the conditional probability for the initial configuration
$(y_1, \cdots, y_N)$ we require
\be\label{init}
P(\mathbf{x};t=0) = \delta_{\mathbf{x},\mathbf{y}}.
\ee
In the following we show that the choice 
\be
\label{s123}
S_{(1,\dots,N)} = \exp\left(-i \sum_{j=1}^N y_j k_{j}\right) \prod_{i=1}^N v_i^{-y_i}
\ee
with an appropriate (see later) definition of the contours of integration in the
complex $k_i$ planes (\ref{invfour}) indeed reproduces the required initial condition.

In order to write the conditional probability in a compact form
we define the functions
\begin{equation}\label{function}
F_{k,l}(x,t)=
\frac1{2\pi i}\oint e^{\frac{t}{z}} z^{x-1} 
\prod_{i=1}^{l-1} (1-v_iz)^{-1} \prod_{i=1}^{k-1} (1-v_iz) dz.
\end{equation}
Here, as throughout this paper, the empty product is defined as unity.
The integral has to be taken along a circle of radius $\epsilon$ around the 
origin of the complex plane. Notice that if all the rates are less than a given 
number $v_\text{max}$ (as we assume here) then the integral can be taken just as 
well along the circle with radius $1/v_\text{max}-0$. 
Now we exploit the determinant expression (\ref{det}) 
to obtain
\begin{thm}
{\em(Conditional probability)}
Let $P({\mathbf x}|{\mathbf y};t)$ be the conditional
probability for the TASEP with $N$ particles on the infinite one-dimensional 
lattice with right hopping rates $v_1, v_2, \cdots, v_N$ (from left to right),
starting at sites $y_1, y_2, \cdots, y_N$ at time $t=0$. The conditional
probability of finding these particles on sites $x_1, x_2, \cdots, x_N$ at 
time $t$ can be written as the determinant
\begin{multline}
\label{mainres}
P({\mathbf x}|{\mathbf y};t) = \prod_{i=1}^N \left( e^{-t v_i} 
{v_i}^{x_i-y_i} \right) \\
\times \left|
\begin{array}{cccc}
F_{1,1}(x_1-y_1,t) & F_{1,2}(x_2-y_1,t) & \cdots & F_{1,N}(x_N-y_1,t) \\
F_{2,1}(x_1-y_2,t) & F_{2,2}(x_2-y_2,t) & \cdots & F_{2,N}(x_N-y_2,t) \\
\vdots & \vdots & & \vdots \\
F_{N,1}(x_1-y_N,t) & F_{N,2}(x_2-y_N,t) & \cdots & F_{N,N}(x_N-y_N,t) 
\end{array}
\right|
\end{multline} 
\end{thm}
\begin{remark}
The expression (\ref{mainres}) is a generalization of the result 
found in \cite{Schu97} for the case of equal hopping rates $v_i=1$.
The functions $F_{k,l}(x,t)$ then reduce to $e^t F_{l-k}(x,t)$ of that work.\\
\end{remark}
\begin{proof}
One has to show that $P({\mathbf x}|{\mathbf y};t)$ satisfies
(\ref{bc}), (\ref{me2}) and (\ref{init}). Using the specific form of $P_\mathbf{k}$ given by (\ref{det}) and (\ref{s123}) the inverse Fourier transformation (\ref{invfour}) leads to
\be
\label{intermed}
\prod_{i=1}^N {v_i}^{x_i-y_i} \int \frac{dk_i}{2\pi} e^{-E_kt} \det \left[ \prod_{j=1}^{n-1} (1-v_j e^{ik_m})^{-1} 
\prod_{j=1}^{m-1} (1-v_j e^{ik_m}) e^{i k_m x_n} \right]_{m,n}.
\ee
This expression satisfies (\ref{bc}) and (\ref{me2}) by construction, independently of the contour of integration. Notice that the different rows of the determinant contain different $k$ variables which enables us to perform the integration over $k_i$ separately on each element of row $i$. After changing the variable $k$ to $z=e^{ik}$ (\ref{intermed}) indeed reduces to the r.h.s.\ of (\ref{mainres}). By adopting the argument of \cite{Schu97} one can easily show that the specific choice of the contour of integration in (\ref{function}) indeed yields (\ref{init}).
\end{proof}
In the following $F(x)$ has to be understood as $F(x,t)$.

\begin{cor}
\label{cor1}
Consider the probability 
\[
\tilde P(\mathbf{z}| \mathbf{y}; t) = \Prob(x_1\geq z_1, x_2 \geq z_2, \cdots , x_N \geq z_N| {\mathbf y};t)
\]
that all $N$ particles, starting on $\mathbf{y}$,
have reached ``at least'' the configuration $\mathbf{z}$ where 
$z_1<z_2< \cdots <z_N$. This quantity can be written as a determinant
\begin{multline}
\label{corollary}
\tilde P(\mathbf{z}| \mathbf{y}; t) = 
\prod_{i=1}^N \left( e^{-t v_i} {v_i}^{z_i-y_i} \right) \\
\times \left|
\begin{array}{cccc}
F_{1,2}(z_1-y_1) & F_{1,3}(z_2-y_1) & \cdots & F_{1,N+1}(z_N-y_1) \\
F_{2,2}(z_1-y_2) & F_{2,3}(z_2-y_2) & \cdots & F_{2,N+1}(z_N-y_2) \\
\vdots & \vdots & & \vdots \\
F_{N,2}(z_1-y_N) & F_{N,3}(z_2-y_N) & \cdots & F_{N,N+1}(z_N-y_N) 
\end{array}
\right|.
\end{multline} 
\end{cor}
\begin{proof} 
One has to write $\tilde P(\mathbf{z}|\mathbf{y};t)$ in the form 
\be
\tilde P(\mathbf{z}|\mathbf{y};t)=
\sum_{x_N=z_N}^\infty \sum_{x_{N-1}=z_{N-1}}^{x_N-1} \cdots \sum_{x_3=z_3}^{x_4-1}\sum_{x_2=z_2}^{x_3-1}\sum_{x_1=z_1}^{x_2-1} 
P({\mathbf x},{\mathbf y};t)
\ee
then use (\ref{mainres}). Applying the identities 
\begin{gather}
\label{sum1}
\sum_{x=x_1}^{x_2} {v_l}^x F_{k,l}(x) = 
{v_l}^{x_1}F_{k,l+1}(x_1) - {v_l}^{x_2+1}F_{k,l+1}(x_2+1) \\
\label{sum2}
\sum_{x=x_1}^{x_2} {v_{k-1}}^x F_{k,l}(x) = 
{v_{k-1}}^{x_1}F_{k-1,l}(x_1) - {v_{k-1}}^{x_2+1}F_{k-1,l}(x_2+1)
\end{gather} 
and standard determinant-manipulations then leads to (\ref{corollary}).
The identities (\ref{sum1},\ref{sum2}) follow by straightforward computation from the definitions
(\ref{function}).
\end{proof}

\section{Current distribution}

We consider now the step function initial state where the lattice is
fully occupied up to the lattice site $x=0$ and empty for $x>0$.
The time evolution of the mean density (under Eulerian scaling) has been 
obtained in \cite{Sepp99} for a large class of random distributions of
rates $v_i$. An interesting problem are the fluctuations of the current
which were first obtained in \cite{Joha00} for the homogeneous system
with $v_i=1$. As a step towards characterizing the long-time
properties of these fluctuations (which we do not consider) we derive here for 
finite times an exact expression for the probability that the 
time-integrated current across a given bond (i.e. the number of particles
that have crossed that bond up to time $t$) equals some number $N$.
This quantity is related to the distribution of the distance travelled by the 
first particle with the initial condition where $N$ particles sit on 
consecutive sites. More precisely, 
\begin{thm} {\em (Current distribution)}
choosing as initial sites
the set $\mathbf{y_N}=(1-N,\cdots, 0)$ the probability that {\em all} particles have crossed the bond $x-1,x$ up to time $t$ can be written as 
\begin{multline}
\label{QN}
Q_N(x,t)=
\prod_{i=1}^N \left(e^{-t v_i} {v_i}^{x+N-1} \right) \\ \times 
\left|
\begin{array}{cccc}
F_{1,2}(x+N-1) & F_{1,3}(x+N) &\cdots & F_{1,N+1}(x+2N-2) \\
F_{2,2}(x+N-2) & F_{2,3}(x+N-1) &\cdots & F_{2,N+1}(x+2N-3) \\
\vdots & \vdots &  & \vdots \\
F_{N,2}(x) & F_{N,3}(x+1) &\cdots & F_{N,N+1}(x+N-1) 
\end{array}
\right|.
\end{multline}
\end{thm}
\begin{remark}
In the case of the initial condition $\mathbf{y_\infty}$ when the lattice is fully occupied on the left and empty on the right of the origin $Q_N(x,t)$ is the probability that {\em at least} $N$ particles have crossed the bond $x-1,x$ up to time $t$. On the other hand this is equal to the probability that the $N$th particle (from the right) reached {\em at least} the site $x$ up to time $t$.
\end{remark}
\begin{proof}
Notice that $Q_N(x,t)=\tilde P(x, x+1, \cdots, x+N-1|\mathbf{y_N};t)$. By using (\ref{corollary}) one arrives at (\ref{QN}).
\end{proof}
\begin{remark}
The determinant appearing in (\ref{QN}) has several equivalent forms. One can derive these by using the identities 
\begin{gather}
\label{spec1}
F_{k,l}(x) = F_{k,l+1}(x) - v_l F_{k,l+1}(x+1) \\
\label{spec2}
F_{k,l}(x) = F_{k-1,l}(x) - v_{k-1}F_{k-1,l}(x+1),
\end{gather} 
which are special cases of (\ref{sum1},\ref{sum2}) corresponding to $x_1=x_2$. 
An example is the following
\end{remark}
\begin{multline}\label{alt1}
Q_N(x,t)=
\prod_{i=1}^N \left(e^{-t v_i} {v_i}^{x+N-1} \right) \\
\times \left|
\begin{array}{cccc}
F_{1,2}(x+N-1) & F_{1,3}(x+N) &\cdots & F_{1,N+1}(x+2N-2) \\
F_{1,2}(x+N-2) & F_{1,3}(x+N-1) &\cdots & F_{1,N+1}(x+2N-3) \\
\vdots & \vdots &  & \vdots \\
F_{1,2}(x) & F_{1,3}(x+1) &\cdots & F_{1,N+1}(x+N-1) 
\end{array}
\right|.
\end{multline} 
Another equivalent expression is
\begin{multline}\label{alt2}
Q_N(x,t)=
\prod_{i=1}^N \left(e^{-t v_i} {v_i}^{x+N-1} \right) \\
\times \left|
\begin{array}{ccccc}
F_{1,N+1}(x+N-1) & F_{1,N+1}(x+N) &\cdots & F_{1,N+1}(x+2N-2) \\
F_{1,N+1}(x+N-2) & F_{1,N+1}(x+N-1) &\cdots & F_{1,N+1}(x+2N-3) \\
\vdots & \vdots &  & \vdots \\
F_{1,N+1}(x) & F_{1,N+1}(x+1) &\cdots & F_{1,N+1}(x+N-1) 
\end{array}
\right|.
\end{multline}

\section{Final remarks}

The study of current fluctuations provides insight into universal properties
of far-from-equilibrium systems and therefore has attracted a considerable
amount of attention in recent years 
\cite{Derr98,Joha00,Prae00,Prae02,Derr04,Sasa04,Rako05,Harr05,Ferr04,Ferr05,Sasa05,Sasa05b}.
There is, however, to our knowledge no work on current fluctuations in
particle systems with more than one conservation law. The result (\ref{alt2})
is a first step in this direction.
The form of (\ref{alt2}) indicates that the distribution $Q_N(x,t)$ is independent
of the order of the hopping rates of particles (e.g. $v_1=a, v_2=b, v_3=c$ 
would lead to the same $Q_N(x,t)$ as $v_1=c, v_2=a, v_3=b$) since these hopping 
rates enter symmetrically in $F_{1,N+1}(x,t)$. This at first sight perhaps 
counterintuitive property becomes obvious in the growth presentation
of the exclusion process \cite{Sepp99} and also in the last passage percolation
picture \cite{Joha00,Prae02} where the hopping rates $v_i$ represent line defects with a 
different distribution of random energies. For the
statistical weight of paths crossing these defects it is clearly irrelevant
in which order they occur. In the interpretation of the TASEP with
particle-dependent rates as an $N$-component lattice gas this means that
the nonergodicity of the process implied in the absence of passing
is irrelevant for the fluctuations of the total current.

A next step would be to study the long time behaviour of the current fluctuations. This was calculated recently for the homogeneous ASEP \cite{Joha00,Prae02,Sasa04,Rako05,Sasa05,Ferr05} and the polynuclear growth model (PNG) \cite{Prae00,Ferr04,Sasa05b}. It was found that the current fluctuations in the long time limit are characterized by universal scaling forms, which are given explicitly in terms of Tracy-Widom distributions known from random matrix theory. It would be of interest to see whether this universality is preserved in the case of particle-wise disorder.    

Recently Priezzhev \cite{Prie03} has extended the Bethe ansatz solution for 
the usual TASEP on $\mathbb{Z}$
to periodic boundary conditions. It would be interesting to generalize this 
analysis to the present case. It would also be interesting to investigate
whether determinant solutions can be obtained for integrable models with 
passing \cite{Kari99a,Kari99b,Khor00,Alim02}.

\section*{Acknowledgments}
A.R. acknowledges financial support by the Deutsche Forschungsgemeinschaft.

\end{document}